\documentclass[aps,prd,preprintnumbers,superscriptaddress,showpacs,nofootinbib,11pt,floatfix]{revtex4}
\usepackage{graphicx}
\usepackage{dcolumn}
\usepackage{bm}
\usepackage{amsmath}
\usepackage{amssymb}
\def\mhc{m_{H^\pm}}
\def\dhch{\Delta_{(H^\pm,\, h^0)}}
\def\dhcw{\Delta_{(H^\pm,\, W)}}
\def\dhw{\Delta_{(h^0,\, W)}}
\def\dtb{\Delta_{(u_i,\, d_j)}}
\def\dec{H^+\to W^+\gamma}

\def\beq{\begin{equation}}
\def\eeq{\end{equation}}
\def\[{\left[}
\def\]{\right]}

\def\gsim{\lower.7ex\hbox{$\;\stackrel{\textstyle>}{\sim}\;$}}
\def\lsim{\lower.7ex\hbox{$\;\stackrel{\textstyle<}{\sim}\;$}}

\begin{document}
\begin{flushright}
{DCP-10-02}\\
\end{flushright}
\title{Implications of  Yukawa Textures in the decay $H^+ \to W^+ \gamma$ within the 2HDM-III}
\author{J. E. Barradas-Guevara}
\email{barradas@fcfm.buap.mx} \affiliation{Fac. de Cs.
F\'{\i}sico-Matem\'aticas, BUAP. Apdo. Postal 1364, C.P. 72000
Puebla, Pue., M\'exico.}
\author{F. C\'azarez-Bush}
\email{federico@matem.unam.mx} \affiliation{Instituto de Matem\'aticas, Universidad Nacional Aut\'onoma de M\'exico,
Cd. Universitaria, M\'exico, D.F. 01000, M\'exico.}
\author{A. Cordero-Cid}
\email{acordero@ece.buap.mx} \affiliation{Facultad de Ciencias de la
Electr\'onica, Benem\'erita Universidad Aut\'onoma de Puebla, Apdo. Postal 542, C. P. 72570 Puebla, Pue.,
M\'exico.}
\author{O. F\'elix-Beltr\'an}
\email{olga_flix@@ece.buap.mx} \affiliation{Facultad de Ciencias de la
Electr\'onica, Benem\'erita Universidad Aut\'onoma de Puebla, Apdo. Postal 542, C. P. 72570 Puebla, Pue.,
M\'exico.}
\author{J. Hern\' andez-S\' anchez}
\email{jaimeh@ece.buap.mx} \affiliation{Facultad de Ciencias de la
Electr\'onica, Benem\'erita Universidad Aut\'onoma de Puebla, Apdo. Postal 542, C. P. 72570 Puebla, Pue.,
M\'exico.}
\affiliation{Dual C-P Institute of High Energy Physics, Puebla, Pue., M\'exico.}
\author{R. Noriega-Papaqui}
\email{rnoriega@uaeh.edu.mx}
\affiliation{Centro de Investigaci\'on en Matem\'aticas,
Universidad Aut\'onoma del Estado de Hidalgo,
Carr. Pachuca-Tulancingo Km. 4.5, C.P. 42184, Pachuca, Hgo., M\'exico.}
\affiliation{Dual C-P Institute of High Energy Physics, Puebla, Pue., M\'exico.}
\date{\today}
\begin{abstract}
\noindent
We discuss the implications of assuming a four-zero Yukawa texture
for  the charged Higgs  decay $H^+ \to W^+ \gamma$ at one-loop level, within the context of
the general 2-Higgs Doublet Model of Type III. We begin by presenting a
detailed analysis of the charged Higgs boson
couplings with heavy quarks and the resulting effects on its
decays. In particular, we present the possible  enhancement of the decay  $H^+ \to W^+ \gamma$, whose branching ratio could be of the order  $\sim 10^{-1}$, for charged Higgs mass  of the order 180 GeV. These parameters can still avoid the $B \to X_s \gamma$ constraint and the perturbativity bound. The production of charged Higgs bosons is also sensitive to
the modifications of its couplings, and we evaluate the events rates  at the LHC, including the `direct'
$c\bar{b}\to H^++c.~c.$ and `indirect' $q\bar q, \, gg \to \bar t b H^++c.~c.$
production channels.  We get  an event rate of order 60 (120) for `direct' (`indirect') production reaction,  and including decay $H^+ \to W^+ \gamma $  in the final state, with  an integrated luminosity of $10^5$ $pb^{-1}$.
\end{abstract}
\pacs{12.60.Cn,12.60.Fr,11.30.Er}
\maketitle

\newpage

\section{Introduction}

Detecting a charged Higgs boson during the imminent Large Hadron
Collider (LHC) experimental running would constitute a clear
evidence of physics beyond the Standard Model (SM) \cite{stanmod}.
Charged Higgs bosons appear in many well motivated extensions of the
SM, whose phenomenology has been widely studied over the years
\cite{kanehunt,susyhix1,susyhix2}. In particular, 2-Higgs Doublet
Models (2HDMs), in both Supersymmetry (SUSY) and non-SUSY versions
\cite{LorenzoDiazCruz:2008zz,Djouadi:2005gj}, can be considered as a
prototype of a Higgs sector that includes a charged Higgs boson
($H^\pm$). It is expected that the LHC will allow us to test the
mechanism of Electro-Weak Symmetry Breaking (EWSB) and, in
particular, to probe the properties of charged Higgs bosons, which
represent a unique probe of a weakly-interacting theory, as is the
case of the Minimal Supersymmetric Standard Model (MSSM)
\cite{LorenzoDiazCruz:2008zz} and general 2HDMs of Type I, II, III
and IV (2HDM-I, 2HDM-II, 2HDM-III and 2HDM-IV) \cite{Barger:1989fj},
 or whether strongly-interacting
scenarios are instead realized, like in the old Technicolor models or
the ones discussed
more recently \cite{stronghix}. Ultimately, while many analyses in
this direction can be carried out at the LHC, it will be a future
International Linear Collider (ILC) \cite{ILC} or Compact Linear Collider (CLIC)
\cite{CLIC}
which will have the definite word about exactly which mechanism of mass generation
and which realization of it occurs in Nature.

The 2HDM-II has been quite attractive to date, in part because it coincides
with the Higgs sector of the MSSM, wherein each Higgs doublet couples
to the $u$- or $d$-type fermions separately\footnote{Notice that there exist
significant
differences between the 2HDM-II and MSSM though, when it comes to their
mass/coupling configurations and possible Higgs signals \cite{Kanemura:2009mk}.}.
However, this is only valid at
tree-level \cite{Babu-Kolda}. When radiative effects are included, it
turns out that the MSSM Higgs sector corresponds to the most general
version of the 2HDM, namely the 2HDM-III, whereby both Higgs fields
couple to both quarks and leptons. Thus, we can consider the
2HDM-III as a generic description of physics at a higher scale (of
order TeV or maybe even higher), whose low energy imprints are
reflected in the Yukawa coupling structure. With this idea in
mind,  a detailed study of the 2HDM-III Yukawa
Lagrangian was presented in Refs.\cite{ourthdm3a, DiazCruz:2009ek}, under the assumption of a specific
texture pattern \cite{Fritzsch:2002ga}, which generalizes the
original model of Ref.~\cite{cheng-sher}. Phenomenological
implications of this model for the neutral Higgs sector, including
Lepton Flavour Violation (LFV) and/or
Flavour Changing Neutral Currents (FCNCs) have been presented in a
previous work
\cite{ourthdm3b}. The extension of  such an approach
to investigate charged Higgs boson phenomenology was conducted in Ref.  \cite{DiazCruz:2009ek},  which discussed the implications of this Yukawa texture for the charged Higgs
boson properties (masses and couplings) and  the resulting pattern of
charged Higgs boson decays and main production reactions at the LHC.

Decays of charged Higgs bosons have been studied in the literature,
including the radiative modes $W^{\pm}\gamma, W^{\pm}Z^0$
\cite{hcdecay, HernandezSanchez:2004tq}, mostly within the context of the 2HDM-II or its SUSY
incarnation (i.e., the MSSM), but also by
using an effective Lagrangian extension of the
2HDM \cite{ourpaper}. More recently, within an extension of the
MSSM with one Complex Higgs Triplet (MSSM+1CHT) \cite{ourtriplets,
Barradas-Guevara:2004qi}. While the decay mode $H^+\to
W^+\gamma$ is forbidden at the tree level due to electromagnetic
gauge invariance, the $H^+\to W^+Z$ decay can be induced at this
order in models including Higgs triplets or more complicated
representations \cite{kanehunt,ourtriplets}. In spite of their suppressed branching
ratios, these decay modes are very interesting,  due to
the fact that their experimental study may provide important
information concerning the underlying structure of the gauge and
scalar sectors; these channels have a clear signature and
might be at the reach of current and future particle colliders. 
Charged Higgs boson production at hadron
colliders was studied long ago \cite{mapaetal,diaz-sampayo} and, more recently,
systematic calculations of production processes at the LHC
have been presented \cite{newhcprod}.

Current bounds on the mass of a charged Higgs boson have been obtained at
Tevatron, by studying the top decay $t \to b \, H^+$, which already
eliminates large regions of the parameter space
\cite{Abulencia:2005jd}, whereas LEP2 bounds imply that, approximately,
$m_{H^{+}} > 80$ GeV \cite{lepbounds,partdat}, rather
model independently.
Concerning theoretical limits,
tree-level unitarity bounds on the 2HDM Higgs masses have been studied
in generic 2HDMs and in particular an upper limit for the charged
Higgs mass of
800 GeV or so can be obtained, according to  the results of
Ref.~\cite{unitarity}.

In this paper we extend previous studies, to include the decay $\dec$. This paper  is organized as follows. In section
II, we discuss the Higgs-Yukawa sector of the 2HDM-III, in particular,
we derive the expressions for the charged Higgs boson couplings to
heavy fermions. Then, in section
III, we review the expressions for the decays $\dec$  at one-loop level 
and numerical results are presented for some 2HDM-III scenarios,
defined for phenomenological purposes. Actual LHC
event rates for  the main
production mechanisms at the LHC are given in section IV. These
include the $s$-channel
production of charged Higgs bosons through
$c\bar{b}(\bar{c}b)$-fusion \cite{He:1998ie} and the multi-body
more  $q \bar{q}, \, gg \to t \bar{b} H^- + $ c.c. (charge conjugated). These
mechanisms depend crucially on the parameters of the underlying model and
large deviations could be expected in the 2HDM-III
with respect to the 2HDM-II.  Finally, we summarize our results
and present the conclusions in section V. 

\section{The Charged Higgs boson Lagrangian and the fermionic couplings}

We shall follow Refs.~\cite{ourthdm3a, ourthdm3b}, where a
specific four-zero texture has been implemented for the Yukawa
matrices within the 2HDM-III. This allows one to express the couplings
of the neutral and charged Higgs bosons in terms of the fermion
masses, Cabibbo-Kobayashi-Maskawa
(CKM) mixing angles and certain dimensionless parameters,
which are to be bounded by current experimental constraints. Thus,
in order to derive the interactions of the charged Higgs boson, the
Yukawa Lagrangian is written as follows: \beq {\cal{L}}_{Y} =
Y^{u}_1\bar{Q}_L {\tilde \Phi_{1}} u_{R} +
                   Y^{u}_2 \bar{Q}_L {\tilde \Phi_{2}} u_{R} +
Y^{d}_1\bar{Q}_L \Phi_{1} d_{R} + Y^{d}_2 \bar{Q}_L\Phi_{2}d_{R},
\label{lagquarks} \eeq \noindent where $\Phi_{1,2}=(\phi^+_{1,2},
\phi^0_{1,2})^T$ refer to the two Higgs doublets, ${\tilde
\Phi_{1,2}}=i \sigma_{2}\Phi_{1,2}^* $, $Q_{L}$ denotes the
left-handed fermion doublet, $u_{R} $ and $d_{R}$ are the
right-handed fermions singlets and, finally, $Y_{1,2}^{u,d}$ denote the
$(3 \times 3)$ Yukawa matrices. Similarly, one can
write the corresponding Lagrangian for leptons.

After spontaneous EWSB  and including the
diagonalizing matrices for quarks and Higgs bosons\footnote{The
details of both diagonalizations are presented in
Ref.~\cite{ourthdm3a}.}, the interactions of the charged Higgs boson
$H^+$ with quark pairs have the following form:
\begin{eqnarray}
\label{QQH} {\cal{L}}^{\bar{q}_i q_j H^+} & = & \frac{g}{2
\sqrt{2} M_W}  \sum^3_{l=1}\bar{u}_i \left\{  (V_{\rm CKM})_{il} \left[ \tan \beta
\, m_{d_{l}} \, \delta_{lj} -\sec \beta  \left(\frac{\sqrt{2}
M_W}{g}\right) \left( \tilde{Y}^d_2\right)_{lj}  \right] \right.
\nonumber \\
& & + \left[ \cot \beta \, m_{u_{i}} \, \delta_{il} -\csc \beta
\left(\frac{\sqrt{2} M_W}{g}\right) \left(
\tilde{Y}^u_1\right)_{il}^{\dagger}  \right]
(V_{\rm CKM})_{lj} \nonumber \\
& & + (V_{\rm CKM})_{il} \left[ \tan \beta \, m_{d_{l}} \, \delta_{lj}
-\sec \beta  \left(\frac{\sqrt{2} M_W}{g}\right) \left(
\tilde{Y}^d_2\right)_{lj}  \right] \gamma^{5}\\
& & - \left. \left[ \cot \beta \, m_{u_{i}} \, \delta_{il} -\csc
\beta  \left(\frac{\sqrt{2} M_W}{g}\right) \left(
\tilde{Y}^u_1\right)_{il}^{\dagger}  \right] (V_{\rm CKM})_{lj} \,
\gamma^{5} \right\} \, d_{j} \, H^{+},   \nonumber
\end{eqnarray}
where $V_{\rm CKM}$ denotes the mixing matrices of the quark sector (and
similarly for the leptons). The term proportional to $\delta_{ij}$
corresponds to the contribution that would arise within the 2HDM-II,
while the terms proportional to $\tilde{Y}_2^d$ and $\tilde{Y}_1^u$
denote the new contributions from the 2HDM-III. These contributions,
depend on the rotated matrices: $\tilde{Y}_n^{q} = O_q^{T}\, P_q\,
Y^{q}_n \, P_q^\dagger \, O_q$ ($n=1$ when $q=u$, and $n=2$ when
$q=d$ ), where $O_q$ is the diagonalizing matrix, while $P_q$
includes the phases of the Yukawa matrix. In order to evaluate
$\tilde{Y}^{q}_n$ we shall consider that all Yukawa matrices have
the four-Hermitic-texture form~\cite{Fritzsch:2002ga}, and the quark
masses have the same form, which are given by:
\begin{equation} M^q= \left( \begin{array}{ccc}
0 & C_{q} & 0 \\
C_{q}^* & \tilde{B}_{q} & B_{q} \\
0 & B_{q}^*  & A_{q}
\end{array}\right)  \qquad
(q = u, d) .
\end{equation}
To diagonalize these matrices, we use the matrices $O_q$ and $P_q$, in the
following way \cite{Fritzsch:2002ga}:
\begin{equation}
\bar{M}^q = O^T_q \, P_q \, M^{q} \,P^{\dagger}_q \, O_q.
\label{masa-diagonal}
\end{equation}
Then, one can derive a better approximation for the product
$O_q^{T}\, P_q\, Y^{q}_n \, P_q^\dagger \, O_q$, expressing the
rotated matrix $\tilde {Y}^q_n$, in the form
\begin{equation}
\left[ \tilde{Y}_n^{q} \right]_{ij}
= \frac{\sqrt{m^q_i m^q_j}}{v} \, \left[\tilde{\chi}_{n}^q \right]_{ij}
=\frac{\sqrt{m^q_i m^q_j}}{v}\,\left[\chi_{n}^q \right]_{ij}  \, e^{i \vartheta^q_{ij}}.
\end{equation}
In order to perform our phenomenological study, we find it convenient
to rewrite the Lagrangian given in Eq.~(\ref{QQH})  in terms of the
coefficients $ \left[\tilde{\chi}_{n}^q \right]_{ij}$, as follows:
\begin{eqnarray}
\label{LCCH}
{\cal{L}}^{q} & = &
\frac{g}{2 \sqrt{2} M_W} \sum^3_{l=1} \bar{u}_i \left\{  (V_{\rm CKM})_{il} \left[ \tan \beta \, m_{d_{l}} \, \delta_{lj}
-\frac{\sec \beta}{\sqrt{2} }  \,\sqrt{m_{d_l} m_{d_j} } \, \tilde{\chi}^d_{lj}  \right] \right.
\nonumber \\
& & + \left[ \cot \beta \, m_{u_{i}} \, \delta_{il}
  -\frac{\csc \beta}{\sqrt{2} }  \,\sqrt{m_{u_i} m_{u_l} } \, \tilde{\chi}^u_{il} \right]
  (V_{\rm CKM})_{lj} \nonumber \\
& & + (V_{\rm CKM})_{il} \left[ \tan \beta \, m_{d_{l}} \, \delta_{lj}
-\frac{\sec \beta}{\sqrt{2} }  \,\sqrt{m_{d_l} m_{d_j} } \, \tilde{\chi}^d_{lj}   \right] \gamma^{5}
 \\
& & - \left. \left[ \cot \beta \, m_{u_{i}} \, \delta_{il}
  -\frac{\csc \beta}{\sqrt{2} }  \,\sqrt{m_{u_i} m_{u_l} } \, \tilde{\chi}^u_{il}  \right]
  (V_{\rm CKM})_{lj} \, \gamma^{5} \right\} \, d_{j} \, H^{+},   \nonumber
\end{eqnarray}
where we have redefined $\left[ \tilde{\chi}_{1}^u \right]_{ij} =
\tilde{\chi}^u_{ij}$ and $\left[ \tilde{\chi}_{2}^d \right]_{ij} =
\tilde{\chi}^d_{ij}$. Then, from Eq.~(\ref{LCCH}), the couplings $\bar{u}_i d_j H^+$ and $u_i \bar{d}_j
H^-$ are given by:
\begin{eqnarray}
\label{coups1} g_{H^+\bar{u_i}d_j} &=& -\frac{ig}{ 2\sqrt{2} M_W}
(S_{i j} +P_{i j} \gamma_5), \quad g_{H^- u_i \bar{d_j}}= -\frac{ig
}{2\sqrt{2} M_W}  (S_{i j} -P_{i j} \gamma_5),
\end{eqnarray}
where $S_{i j}$ and $P_{i j}$ are defined as:
\begin{eqnarray}
\label{sp}
S_{i j} & = &  \sum^3_{l=1} \bigg\{ (V_{\rm CKM})_{il} \bigg[ \tan \beta \, m_{d_{l}} \,
\delta_{lj} -\frac{\sec \beta}{\sqrt{2} }  \,\sqrt{m_{d_l} m_{d_j} }
\, \tilde{\chi}^d_{lj}  \bigg]
\nonumber \\
& & + \bigg[ \cot \beta \, m_{u_{i}} \, \delta_{il}
  -\frac{\csc \beta}{\sqrt{2} }  \,\sqrt{m_{u_i} m_{u_l} } \, \tilde{\chi}^u_{il} \bigg]
  (V_{\rm CKM})_{lj} \bigg\}, \nonumber \\
P_{i j} & = & \sum^3_{l=1}  \bigg\{ (V_{\rm CKM})_{il} \bigg[ \tan \beta \, m_{d_{l}} \,
\delta_{lj}
-\frac{\sec \beta}{\sqrt{2} }  \,\sqrt{m_{d_l} m_{d_j} } \, \tilde{\chi}^d_{lj}   \bigg]  \\
& & - \bigg[ \cot \beta \, m_{u_{i}} \, \delta_{il}
  -\frac{\csc \beta}{\sqrt{2} }  \,\sqrt{m_{u_i} m_{u_l} } \, \tilde{\chi}^u_{il}  \bigg]
  (V_{\rm CKM})_{lj} \bigg\}. \nonumber
\end{eqnarray}
As it was discussed in Ref.~\cite{ourthdm3a}, most low-energy
processes imply weak bounds on the coefficients
$\tilde{\chi}^q_{ij}$, which turn out to be of $O(1)$. However, some
important constraints on $\tan \beta$ have started to appear, based
on $B$-physics \cite{Dudley:2009zi}. In order to discuss these
results we find convenient to generalize the notation of Ref.~\cite{Borzumati:1998nx} and define the couplings
$\bar{u}_i d_j H^+$ and $u_i \bar{d}_j H^-$ in terms of the
matrices $X_{ij}$, $Y_{ij}$ and $Z_{ij}$ (for leptons). In our case
these matrices are given by:
\begin{eqnarray}
X_{l j} & = &   \bigg[ \tan \beta \,
\delta_{lj} -\frac{\sec \beta}{\sqrt{2} }  \,\sqrt{\frac{m_{d_j}}{m_{d_l}}  }
\, \tilde{\chi}^d_{lj}  \bigg],
\nonumber \\
Y_{i l} & = &  \bigg[ \cot \beta \,  \delta_{il}
  -\frac{\csc \beta}{\sqrt{2} }  \,\sqrt{\frac{ m_{u_l}}{m_{u_i}} } \, \tilde{\chi}^u_{il}  \bigg] .
\end{eqnarray}
where $X_{lj}$ and $Y_{il}$ are related with $S_{ij}$ and $P_{ij}$
defined in the Eq.~(\ref{sp}) as follows:
\begin{eqnarray}
S_{i j} & = &  \sum^3_{l=1} (V_{\rm CKM})_{il}  \, m_{d_{l}} \, X_{lj}
 +  m_{u_{i}} \, Y_{il}  (V_{\rm CKM})_{lj}, \nonumber \\
P_{i j} & = &  \sum^3_{l=1} (V_{\rm CKM})_{il}  \, m_{d_{l}} \, X_{lj}
 -  m_{u_{i}} \, Y_{il}  (V_{\rm CKM})_{lj}.
\end{eqnarray}
The 33 elements of these matrices reduce to the expressions for the
parameters X,Y,Z ($=X_{33},Y_{33},Z_{33}$) used in
Ref.\cite{Borzumati:1998nx}. Based on the analysis of $B \to X_s
\gamma$ \cite{Borzumati:1998nx, Xiao:2003ya}, it is claimed that $X
\leq 20$ and $Y \leq 1.7$ for $m_{H^+} > 250$ GeV, while for a
lighter charged Higgs boson mass, $m_{H^+} \sim 180$ GeV, one gets
$(X,Y) \leq (18,0.5)$. In  recently work  we get the values of $(X,Y)$
as a function of $\tan \beta$ within our model. Thus, we find
the bounds: $|\chi_{33}^{u,d}| \lsim 1$ for $0.1 < \tan \beta
\leq 70$ \cite{DiazCruz:2009ek}. Although in our model there are additional contributions
(for instance from $c$-quarks, which are proportional to $X_{23}$),
they are not relevant because the Wilson coefficients in the
analysis of $B \to X_s \gamma$ are functions of $m_{c}^2/M_W^2$ or
$m_{c}^2/m_{H^+}^2$ \cite{Deshpande:1987nr}, that is, negligible
 when compared to the leading $X_{33}$ effects, whose Wilson
coefficients depend on $m_t^2/M_W^2$ or $m_t^2/m_{H^+}^2$. Other
constraints on the charged Higgs mass and $\tan\beta$,  can be obtained the anomalous magnetic momentum of the muon based on
$\Delta a_{\mu}$, the $\rho$ parameter, as well as B-decays into the
tau lepton, can be obtained
\cite{BowserChao:1998yp,WahabElKaffas:2007xd}. For instance, as can
be read from Ref.\cite{Isidori:2007ed}, one has that the decay $B
\to \tau \nu$, implies a constraint such that for $m_{H^+}=200$
(300) GeV, values of $\tan\beta$ less than about 30 (50) are still
allowed, within MSSM or THDM-II: However, these constraints can
only be taken as estimates, as it is likely that they would be
modified for THDM-III. In summary, we find that low energy
constraints still allow to have $\tilde{\chi}^q_{ij}=
O(1)$\footnote{A more detailed analysis that includes the most
recent data is underway \cite{lorenzoetal, Mahmoudi:2009zx}.}.  
On the other hand, the condition $\frac{\Gamma_{H^+}}{m_{H^+}} < \frac{1}{2}$ in the
frame of the 2HDM-II implies $\frac{\Gamma_{H^+}}{m_{H^+}} \approx
\frac{3G_F m_t^2}{4\sqrt{2}\pi\tan\beta^2}$ which leads to $0.3
\lsim \tan\beta \lsim 130$. However, in the 2HDM-III we have that
$\frac{\Gamma_{H^+}}{m_{H^+}} \approx \frac{3G_F
m_t^2}{4\sqrt{2}\pi\tan\beta^2} \bigg(
\frac{1}{1-\frac{\tilde{\chi}^u_{33}}{\sqrt{2} \cos\beta}}\bigg)^2$,
we have checked numerically that this leads to $0.08 < \tan\beta <
200$ when $|\tilde{\chi}^u_{33}| \approx 1$ and  $0.3 < \tan\beta <
130$ as long as $|\tilde{\chi}^u_{33}| \to 0$ recovering the result
for the case of the 2HDM-II \cite{Barger:1989fj, Chankowski:1999ta}.
In this sense, if we consider the constraints imposed by the
perturbativity bound, a portion of the low $\tan\beta$ appearing in
some graphs would be excluded. However, we have decided to keep that
range both to show the behaviour of the quantities of interest, and
also  to keep in mind that such criteria
(perturbativity) should be taken as an order of magnitude
constraint.

\section{\bf Decay  $H^+ \to W^+ \gamma$}

In the calculation of the width  for decay $H^+ \to W^+ \gamma$, we employe a nonlinear
$R_\xi$-gauge, which leads to considerably simplifications due to
the fact that some unphysical vertices are removed from the
interaction Lagrangian \cite{HernandezSanchez:2004tq}. We have shown that such a gauge not only
reduces considerably the number of Feynman diagrams but also
renders manifestly gauge-invariant and ultraviolet-finite
amplitudes. Apart from emphasizing the advantages of using the
nonlinear $R_\xi$-gauge, we will analyze the $\dec$ decay in some
scenarios which are still consistent with the most recent bounds
discussed in the previous section. 

\subsection{The gauge fixing procedure}
We now would like to comment on the gauge-fixing procedure which
was used to simplify our calculation. To this end we introduce the
following gauge-fixing functions \cite{Fujikawa}:
\begin{eqnarray}
f^+&=&\left(D^e_\mu +\frac{igs^2_W}{c_W}Z_\mu \right)W^{+\mu}-i\xi m_WG^+_W, \\
f^Z&=&\partial_\mu Z^\mu-\xi m_ZG_Z,\\
f^A&=&\partial_\mu A^\mu,
\end{eqnarray}
\noindent with $D^e_\mu$ the electromagnetic covariant derivative
and $\xi$ the gauge parameter. Note that $f^+$ is nonlinear and
transforms covariantly under the electromagnetic gauge group. This
gauge-fixing procedure is suited to remove the unphysical vertices
$WG_W\gamma$ and $WG_WZ$, which arise in the Higgs kinetic-energy
sector, and also modifies the Yang-Mills sector. One important
result is that the expression for the $WW\gamma$ vertex satisfies
a QED-like Ward identity, which turns out to be very useful in
loop calculations. In particular, in the Feynman-t'Hooft gauge the
Feynman rule for the $W^+_\rho(q)W^-_\nu (p) A_\mu (k)$ coupling
can be written as
\begin{equation}
\Gamma^{WW\gamma}_{\rho \nu \mu}=-ie\left(g_{\mu
\nu}(p-k+q)_\rho+g_{\nu \rho}(q-p)_\mu+g_{\mu
\rho}(k-p-q)_\nu\right),
\end{equation}
where all the momenta are incoming. It is easy to see that this
expression satisfies a QED-like Ward identity.
The remaining Feynman rules necessary for our calculation do not
depend on the gauge fixing procedure.  As far as the Yukawa
couplings are concerned, we will concentrate on the type-III 2HDM.

\subsection{Decays of the charged Higgs boson at tree level}

 The
expressions for the charged Higgs boson decay widths $H^+ \to u_i
\bar{d}_j$ are of the form:
\begin{eqnarray}
\Gamma (H^+ \to u_i \bar{d}_j) &=& \frac{3 g^2}{32 \pi M_W^2 m_{H^+}^3 } \lambda^{1/2}(m_{H^+}^2, m_{u_i}^2, m_{d_j}^2) \nonumber\\
& & \times \, \bigg( \frac{1}{2}\bigg[ m_{H^+}^2-m_{u_i}^2-m_{d_j}^2\bigg] (S_{ij}^2+P_{ij}^2)- m_{u_i} m_{d_j}
(S_{ij}^2- P_{ij}^2 ) \bigg),
\end{eqnarray}
where $\lambda$ is the usual kinematic factor $\lambda(a,b,c)=
(a-b-c)^2-4bc$. When we replace $\tilde{\chi}_{ud} \to 0$, the
formulae of the decays width become those of the 2HDM-II: see, e.g.,
Ref.~\cite{kanehunt}. Furthermore, the expressions for the charged Higgs
boson decay widths of the bosonic modes  remain the same as in the
2HDM-II.  Recently \cite{DiazCruz:2009ek}, we have studied that the effect of
the modified Higgs couplings typical of the 2HDM-III
 shows up clearly in the pattern of charged Higgs
boson decays, which can be very different from the 2HDM-II case and thus
 enrich the possibilities to search for
$H^{\pm}$ states at current (Tevatron) and future (LHC, ILC/CLIC) machines.

\subsection{Decay $\dec$ induced at one-loop level}

\begin{figure}[!hbt]
\centering
\includegraphics[width=3.in]{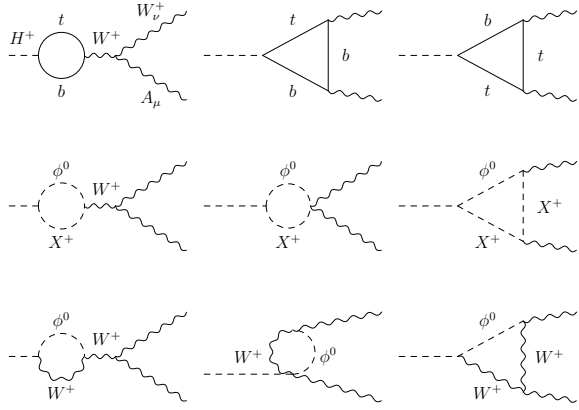}
\caption{\label{hwgdiag}Feynman diagrams contributing to the
$H^+\to W^{+}\gamma$ decay in the nonlinear $R_\xi$-gauge.
$\phi^0$ stands for $h^0$ and $H^0$, and $X^+$ for $H^+$ and
$G_W^+$. }
\end{figure}

In the nonlinear $R_\xi$-gauge, the decay $H^+\to W^+\gamma$
receives contributions from the Feynman diagrams shown in Fig.
\ref{hwgdiag}. As far as the fermionic sector is concerned, the
main contribution comes from the third-generation quarks $(t,b)$, which
induce three diagrams. However, we also considered the quarks contribution $(c,b)$ and $(t,s)$ because their coupling with the charged Higgs can be important \cite{DiazCruz:2009ek}, which induce six more diagram.   Whereas in the bosonic sector there are
contributions from the ($H^\pm,\,\phi^0$) and ($W^\pm,\,\phi^0$)
pairs, with $\phi^0=h^0$ or $H^0$. We would like to emphasize that
the nonlinear $R_\xi$-gauge considerably simplifies the
calculation of the decay $\dec$. First of all, the removal of the
unphysical vertex $W^\pm G^\mp_W \gamma$, and the tadpole graphs  vanish \cite{HernandezSanchez:2004tq}. Apart from these simplifications, we will be able to group the Feynman diagrams into subsets which separately
yield a manifestly gauge-invariant amplitude free of ultraviolet
singularities.
Once the amplitude for every Feynman diagram shown in Fig. \ref{hwgdiag} was written down, the Passarino-Veltman method
\cite{Passarino} was applied to express it in terms of scalar
integrals, which are suitable for numerical evaluation \cite{FF}.
The full amplitude for the decay $H^+(p)\to W^+(q) \gamma (k)$ can
be written as
\begin{equation}
\mathcal{M} ( H^+ \to W^+ \gamma  ) =
 \frac{\alpha^{3/2}}{2 \sqrt{\pi} m_W s_w^2}
 \left( G\,\left(q\cdot k \;g_{\mu\nu}-
  k_{\nu} q_\mu\right) - i\, H\, \epsilon_{\mu \nu \alpha \beta}
k^{\alpha}  q^\beta \right)\epsilon^{\mu } (k) \epsilon^{\nu }
(q), \label{amplitude}
\end{equation}
\noindent which is manifestly gauge invariant. The $H$ function
only receives contributions from the quarks:
\begin{equation}
H = \frac{i}{\dhcw} \sum_{i,j}^3 \left(3\, \lambda_{ij} \Delta
B_{(H^\pm,W,d_j,u_i)}+\dhcw\left( \lambda^d_{ij} \, C_{(d_j,u_i)}-2\, \lambda_{ij}^u\,
C_{(u_i,d_j)}\right)\right),
\end{equation}
where
\begin{equation}\lambda_{ij}=-\frac{1}{2}\bigg( m_{u_i} (P_{ij}+S_{ij})- m_{d_j} (P_{ij}-S_{ij}) \bigg),
\end{equation}
\begin{equation}
\lambda_{ij}^u=\frac{1}{2} m_{u_i} (P_{ij}+S_{ij}),
\end{equation}
\begin{equation}
\lambda_{ij}^d=\frac{1}{2} m_{d_i} (P_{ij}-S_{ij}),
\end{equation}
\noindent here, we have introduced the shorthand notation
$\Delta_{(a,\,b)}\equiv m_a^2-m_b^2$, $\Delta
B_{(a,b,c,d)}=B_0\left(m_a^2,m_c^2,m_d^2\right)-
B_0\left(m_b^2,m_c^2,m_d^2\right)$,\footnote{When $a=0$, $m_a$
must be substituted by 0.} and $C_{(a,b)}\equiv
C_0\left(m^2_{H^\pm},m_W^2,0,m_a^2,m_b^2,m_a^2\right)$, where
$B_0$ and $C_0$ stand for Passarino-Veltman scalar integrals. From the
$\Delta B_{(a,b,c,d)}$ definition, it is clear that the $H$
function is ultraviolet finite. Again, when  we replace $\tilde{\chi}_{ud} \to 0$, the
formulae of the decays width become those of the 2HDM-II \cite{HernandezSanchez:2004tq}.
As for $G$, it can be written as
\begin{eqnarray}
G=\frac{1}{2\,\mhc^2\, \dhcw^2}\left(G_{u_i d_j} +\sum_{\phi^0=H^0,\,
h^0}\left(G_{ H^\pm\phi^0}+ G_{G_W^\pm \phi^0}+ G_{W^\pm
\phi^0}\right)\right), \label{Gdef}
\end{eqnarray}
\noindent where $G_{AB}$ stands for the contribution of the
$(A,\,B)$ pair:
\begin{eqnarray}
G_{u_i d_j} & = & \sum_{ij}^3
\mhc^2 \bigg( \left(    3\,\dtb+m_W^2   \right) \lambda_{ij} + 3 \sigma_{ij} \dhcw \bigg)
\Delta B_{(H^\pm,W,d_j,u_i)} \nonumber \\ &-& 3\,\dtb\dhcw \lambda_{ij} \Delta
B_{(0,H^\pm,d_j,u_i)}  + \mhc^2\dhcw
\bigg( \lambda_{ij}-\big(2\,m_{d_j}^2 \lambda_{ij} -\lambda_{ij}^d\dhcw\big)C_{(d_j,u_i)} \nonumber \\ &+&
\left(2\,m_{u_i}^2 \lambda_{ij}-\lambda_{ij}^u\,\dhcw\right)
C_{(u_i,d_j)}\bigg),
\end{eqnarray}
where
\begin{equation}
\sigma_{ij}=\frac{1}{2}\bigg( m_{u_i} (P_{ij}+S_{ij})+ m_{d_j} (P_{ij}-S_{ij}) \bigg),
\end{equation}
When one takes $\chi_{ij}'s \to 0$, the formulas  for the function $G_{u_i d_j}$ reduce to the 2HDM-II case, see, e. g. \cite{HernandezSanchez:2004tq}.
\begin{eqnarray}
G_{H^\pm h^0 } &=&\lambda_{{h^0 H^\pm}}\Big(\dhch\dhcw \delta
B_{(0,W,H^\pm,h^0)}- \left(2\,\mhc^2\dhch+m_{h^0}^2
m_W^2\right)\delta
B_{(H^\pm,W,H^\pm,h^0)}\nonumber\\&-&\mhc^2\dhcw\left(1+2\,\mhc^2
C_{(H^\pm,h^0)}\right)\Big),
\end{eqnarray}
\begin{eqnarray}
G_{G_W^\pm h^0} &=&\lambda_{{G_W^\pm h^0}}\Big(
\left(m_W^2\dhw+\mhc^2(3m_W^2-2m_{h^0}^2)\right)\delta
B_{(H^\pm,W,h^0,W)}+\dhch\,\dhw \delta
B_{(0,W,h^0,W)}\nonumber\\&+& \mhc^2\dhcw\left(1+2\,m_W^2
C_{(W,h^0)}\right) \Big),
\end{eqnarray}
and
\begin{eqnarray}
G_{W^\pm H^0} & = &\lambda_{W^\pm h^0} \Big(
\left(m_W^2\dhw+\mhc^2\left(7\,m_W^2-2\,m_{h^0}^2-4\,\mhc^2\right)\right)\delta
B_{(H^\pm,W,H^\pm,h^0)} \nonumber \\ &+& \dhcw\dhw \delta
B_{(0,W,H^\pm,h^0)}\nonumber\\&+&\mhc^2\dhcw\left(1-2\,\left(2\,\dhcw-m_W^2\right)C_{(m_W^2,m_{h^0}^2)}\right)
\Big),
\end{eqnarray}
with
\begin{equation}\lambda_{{ H^\pm h^0}}=c_{\alpha-\beta}\left(2\,s_{ \alpha - \beta }\,m_{H^\pm}^2
 +\frac{ c_{\alpha + \beta}}{c^2_{\beta }s^2_\beta}\,\mu_{12}^2
 + \,\left(s_\alpha s_\beta t_\beta - \frac{c_\beta
c_\alpha}{t_\beta}\right)\,m_{h^0}^2 \right),
\end{equation}
\begin{equation}\lambda_{{G_W^\pm h^0}}=\frac{s_{2( \alpha -\beta)}\dhch}{2},\end{equation}
and
\begin{equation}\lambda_{W^\pm h^0}= -\frac{m_W^2 s_{2( \alpha
-\beta)}}{2}.\end{equation}
\noindent Finally, the contribution of the heaviest CP-even scalar
boson $H^0$ is obtained  of the lightest one, once the
substitutions $m_{h^0}\to m_{H^0}$ and $\alpha\to\alpha-\pi/2$ are
done.

It is also
evident that the partial amplitudes induced by the pairs
$(t,\,b)$, $(t,\,s)$, $(c,\,b)$, $(\phi^0,\, W^\pm)$, $(\phi^0,\, H^\pm)$, and
$(\phi^0,\, G_W^\pm)$ are gauge invariant and ultraviolet finite
on their own. Again, this is to be contrasted with the situation
arising in the linear $R_\xi$-gauge, where showing gauge
invariance is somewhat cumbersome, and the cancellation of
ultraviolet divergences in the bosonic sector is achieved only
after adding up all the Feynman diagrams \cite{HernandezSanchez:2004tq}.

Once Eq. (\ref{amplitude}) is squared and the spins of the final
particles are summed over, the decay width can be written as
\begin{equation}
\Gamma(\dec) =  \frac{\alpha^3\,\dhcw^3}{2^7 \pi^2 s_W^4 m^2_W
\mhc^3}  \left( |G|^2+|H|^2\right).
\end{equation}
We will evaluate this decay width for some values of the
parameters of the model.

\subsection{Numerical results and discussion}

\begin{figure}
\centering
\includegraphics[width=6in]{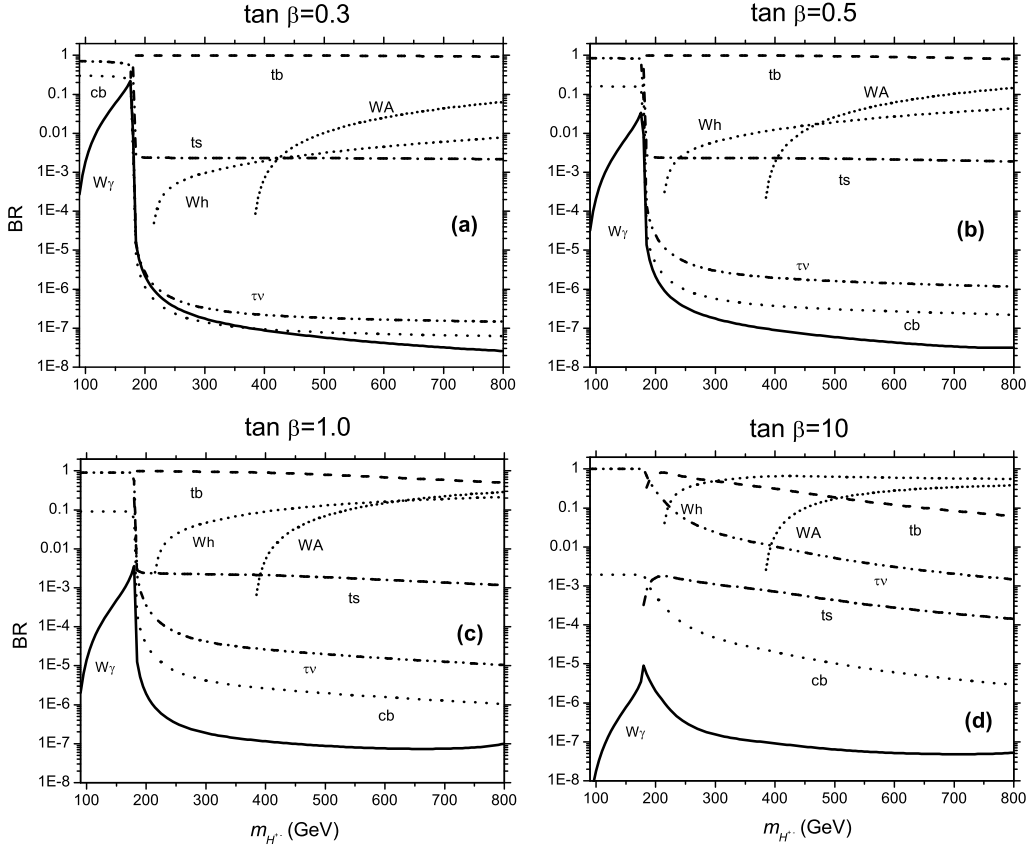}
\caption{The figure shows the BRs of the $H^+$ decaying
into the principal modes in Scenario A, taking
$\tilde{\chi}_{ij}^{u}=-1$, $\tilde{\chi}_{ij}^{d}=-1$, $m_{h^0} =
120$ GeV, $m_{A^0} = 120$ GeV and $\alpha = \pi/2$  for: (a) $\tan
\beta=0.3$, (b) $\tan \beta = 0.5$, (c) $\tan \beta = 1$, (d) $\tan
\beta = 10$. The lines  in each graph correspond to: ($W \gamma$ line) BR($\dec $), ($tb$ line) BR($H^+ \to
t\bar{b} $), ($cb$ line) BR($H^+ \to c\bar{b}$), ($ts $ line) BR($H^+ \to t\bar{s}$),
  ($\tau \nu$ line) BR($H^+ \to \tau^+ \nu_\tau$), (Wh line) BR($H^+ \to W^+ h^0$), (WA line)
BR($H^+ \to W^+ A^0$).} \label{fig:wg}
\end{figure}

Let us now discuss the decay modes of the charged Higgs boson
within our model. Hereafter, we shall refer to two benchmark scenarios,
namely. (i) {\bf Scenario A}:  $\tilde{\chi}^u_{ij}=-1$,
$\tilde{\chi}^d_{ij}=-1$; (ii) {\bf Scenario B}:
$\tilde{\chi}^u_{ij}=1$, $\tilde{\chi}^d_{ij}=1$. We have performed the numerical analysis
of charged Higgs boson decays by taking $\tan\beta=0.3, \, 0.5, \, 1,
\, 10$ and varying the charged Higgs boson mass within the interval
100 GeV $\leq m_{H^{\pm}} \leq$ 800 GeV, further fixing $m_{h^0}= 120$
GeV, $m_{A^0}=300$ GeV and the mixing angle at $\alpha = \pi /2$.
Then the results for the Branching Ratios (BRs) are shown in
Fig. \ref{fig:wg}, and have the following characteristics.

\noindent {\bf Scenario A}. In Fig.~\ref{fig:wg}(a) we present
the BRs for the channels  $\dec$, $H^+ \to t \bar{b}$, $c \bar{b}$, $t
\bar{s}$, $ \tau^+ {\nu_\tau}$, $W^+ h^0$, $W^+ A^0$ as a function
of $m_{H^+}$, for $\tan\beta=0.3$ and fixing $m_{h^0}= 120$ GeV,
$m_{A^0}=300$ GeV and the mixing angle $\alpha = \pi /2$. When
$m_{H^+} < 175$ GeV , we can see that
the dominant decay of the charged Higgs boson is via the mode $\tau \bar{\nu} $, with BR($H_i^+ \to \tau \bar{\nu}$) $\approx 1$, and the same importance we have the mode $c \bar{b}$. For the case 170 GeV $< m_{H^+} <
180$ GeV the mode $t\bar{s}$ is relevant. 
In this range, the mode $W^+ \gamma$ is competitive and becomes of order $10^{-1}$, which it is also very different
from the 2HDM-II and becomes an interesting phenomenological
consequence of the 2HDM-III. We can also observe that, for $m_{H^+}>
180$ GeV, the decay mode $ t \bar{b}$ is dominant (as in the 2HDM-II). In this range mass of the charged Higgs, the channel $\dec$ 
has a BR $\sim 10 ^{-4} $ to $\sim 10 ^{-6} $. 
Now, from
Fig.~\ref{fig:wg}(b), where $\tan \beta  =0.5$, we find that
 the dominant decay mode is into $\tau^+ {\nu_\tau}$ for the
range $m_{H^+}<$ 175 GeV, again for 175 GeV $< m_{H^+} < 180$ GeV
the mode $t\bar{s}$ is the leading one, but for 180
GeV$<m_{H^+}$, the decay channel $t \bar{b}$ becomes relevant,
whereas for the range 170 GeV $<m_{H^+}<$ 180  GeV the mode $W^+ \gamma$ has a BR
of order $10^{-2}$.  Again,   when $m_{H^+} > 180 $ GeV the  $BR(\dec)$ is of order $10^{-4}$ to $10^{-7}$ . Then, see Fig.~\ref{fig:wg}(c), for the
case with $\tan \beta = 1$ one gets that BR($H^+ \to
\tau^+{\nu_\tau}$) $\approx 1$ when $m_{H^+}< 175 $ GeV,  while in the range 175 GeV $< m_{H^+} < 180$ GeV the
mode $t\bar{s}$ is relevant and the $BR(\dec)$ even could be important with an order $10^{-3}$. However for
180 GeV$<m_{H^+}$, the dominant decay of the charged Higgs
boson is the mode  $t \bar{b}$. For $\tan
\beta =10$, we show in plot Fig.~\ref{fig:wg}(d) that the dominant
decay of the charged Higgs boson is the mode $\tau^+ {\nu_\tau}$,
when $m_{H^+}<180$ GeV, but that, for 180 GeV$< m_{H^+} <250$ GeV, the
decay channel $t \bar{b}$ becomes the leading one, whereas for the
range 250 GeV $ < m_{H^+}$, the mode $W^+ h^0$ is relevant. In this case the BR of the decay channel $W^+ \gamma$ is unimportant.

\begin{figure}
\centering
\includegraphics[width=6in]{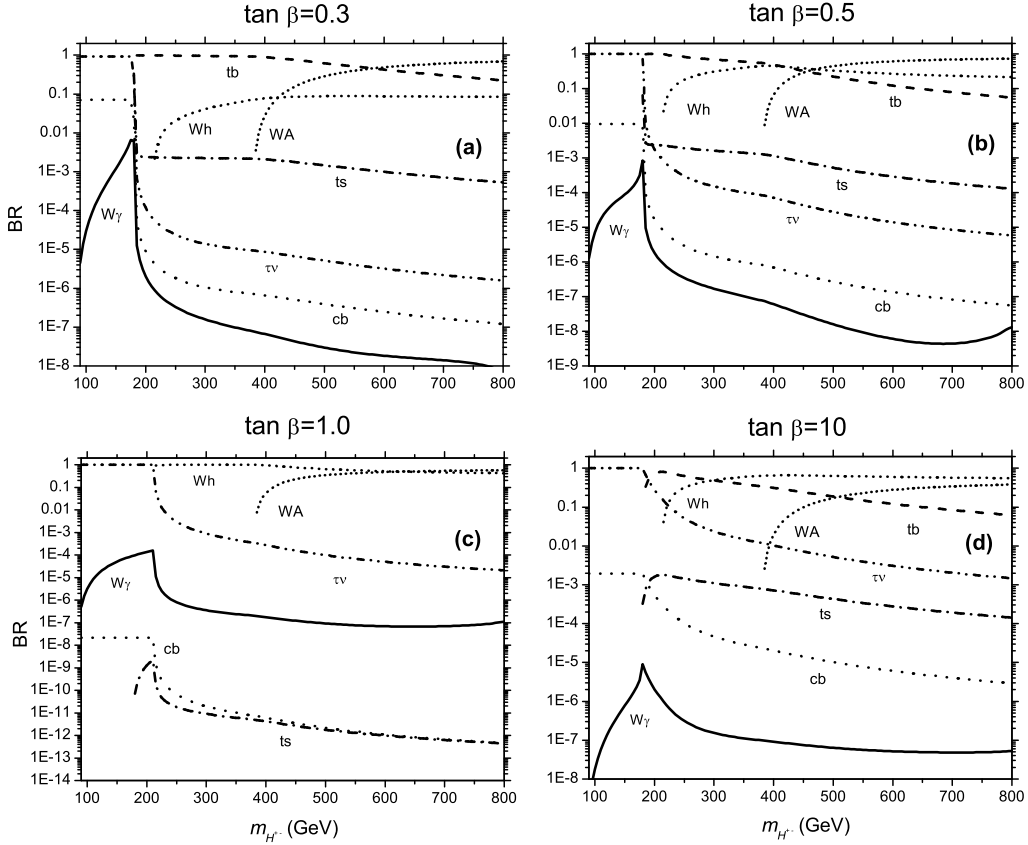}
\caption{The same as in Fig.~\ref{fig:wg} but taking
$\tilde{\chi}_{ij}^{u}=1$, $\tilde{\chi}_{ij}^{d}=1$
(Scenario B).} \label{fig:wga}
\end{figure}
\noindent {\bf Scenario B}. In Fig.~\ref{fig:wga}(a) we present
the BRs for the channels  , $H^+ \to W^+ \gamma, t \bar{b}$, $c \bar{b}$, $t
\bar{s}$, $ \tau^+ {\nu_\tau}$, $W^+ h^0$, $W^+ A^0$ as a function
of $m_{H^+}$, for $\tan\beta=0.3$ and fixing $m_{h^0}= 120$ GeV,
$m_{A^0}=300$ GeV and the mixing angle $\alpha = \pi /2$. When
$m_{H^+} < 175$ GeV , we can see that
the dominant decay of the charged Higgs boson is via the mode $\tau^+ {\nu_\tau}$, with BR($H_i^+ \to \tau^+ {\nu_\tau}$) $\approx 1$. For the case 175 GeV $< m_{H^+} <
180$ GeV the mode $t\bar{s}$ is dominant, whereas for the range 170 GeV $<m_{H^+}<$ 180  GeV the mode $W^+ \gamma$ has a BR
of order $10^{-2}$ . We can also observe that, for $m_{H^+}>
180$ GeV, the decay mode $ t \bar{b}$ is dominant (as in the 2HDM-II). 
Now, from
Fig.~\ref{fig:wga}(b), where $\tan \beta = 0.5$, we find that
 the dominant decay mode is into $\tau^+ {\nu_\tau}$ for the
range $m_{H^+}<$ 175 GeV, again for 175 GeV $< m_{H^+} < 180$ GeV
the mode $t\bar{s}$ is the leading one, while the decay $\dec$ induced at one-loop level has a BR relatively large of order $10 ^{-2}$. 
However for
180 GeV$<m_{H^+}$, the dominant decay of the charged Higgs
boson is the mode  $t \bar{b}$.
Then, see Fig.~\ref{fig:wga}(c), for the
case with $\tan \beta = 1$ one gets that BR($H^+ \to
\tau^+{\nu_\tau}$) $\approx 1$ when $m_{H^+}< 175 $ GeV,  while in the range 175 GeV $< m_{H^+} < 180$ GeV the
mode $t\bar{s}$ is relevant and the $BR(\dec)$ even is of order $10^{-4}$. 
For 180 GeV$<m_{H^+}<600$ GeV, the decay channel $W^+ h^0$ becomes relevant,
whereas for the range 600 GeV $<m_{H^+}$ the mode $W^+ A^0$ is
dominant. It is convenient to mention that this sub-scenario is
special for the mode $t \bar{b}$, because its decay width is zero at
the tree-level, since the CKM contribution is canceled exactly with
the terms of the four-zero texture implemented for the Yukawa
coupling of the 2HDM-III. 
For $\tan \beta =10$, we show in plot Fig.~\ref{fig:wga}(d) that the dominant
decay of the charged Higgs boson is the mode $\tau^+ {\nu_\tau}$,
when $m_{H^+}<180$ GeV, but  for 180 GeV$< m_{H^+} <250$ GeV, the
decay channel $t \bar{b}$ becomes the leading one, whereas for the
range 250 GeV $ < m_{H^+}$, the mode $W^+ h^0$ is again dominant. The decay $\dec$ has a BR of order $10^{-5}$  to $10^{-7}$, which could be irrelevant channel decay for LHC.

In order to cover further the Higgs sector in our
analysis, it is appropriate to also mention how the previous results
change with $m_{h^0}$, $m_{A^0}$ and $\alpha$.  In general the behavior of the decay modes of the
charged Higgs boson is similar to the cases presented above, except
for the decay channel $Wh^0$, which can be important or unimportant. For $\alpha = 0$, this mode has
BR $< 10 ^ {-3} $ when $\tan \beta$ is large. However, for $\tan
\beta <1 $, it becomes the dominant one. In the case $\alpha = \beta$,
the decay channel $Wh^0$ can be the dominant mode,  with a BR of $O(1)$. \\

\begin{figure}
\centering
\includegraphics[width=6in]{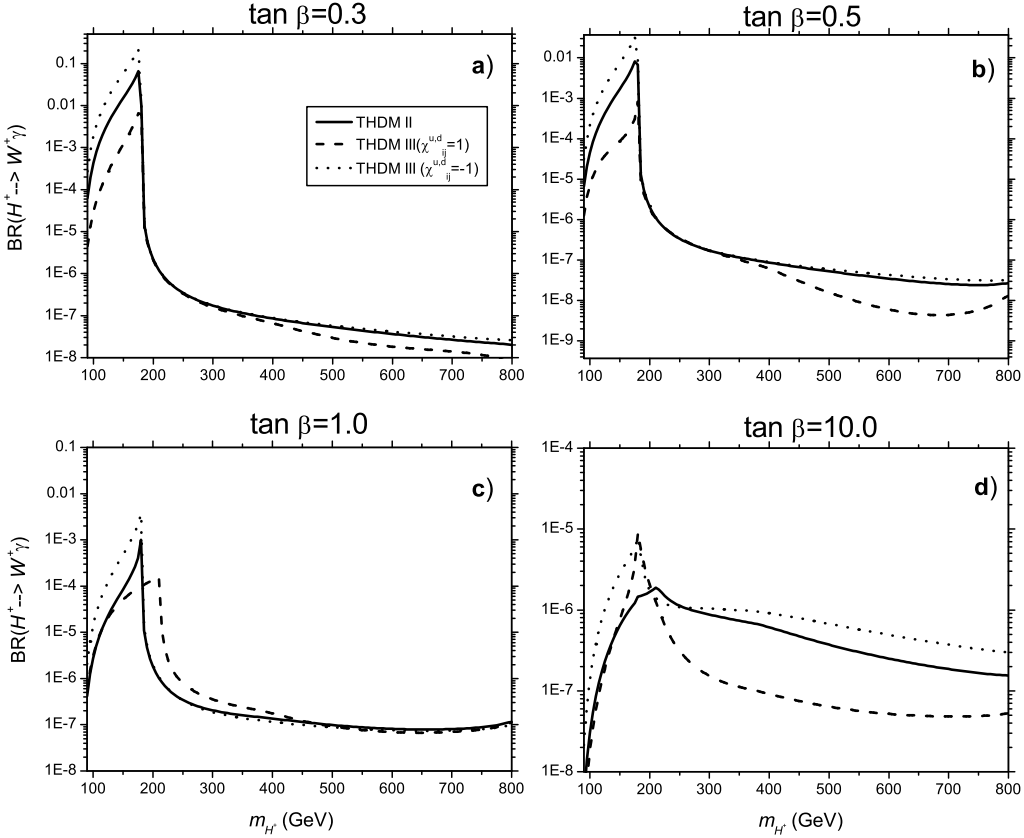}
\caption{The figure shows the BR$(\dec)_{\rm III}, \, {\rm BR} (\dec)_{\rm II} $ {\it vs.} $m_{H^+}$, taking $\tan \beta =0.3, \, 0.5,
\, 1, \, 10 $ for $\tilde{\chi}_{ij}^{u,d}=-1$ (dots), 
$\tilde{\chi}_{ij}^{u,d}=0$  or case 2HDM-II (solid line),  
$\tilde{\chi}_{ij}^{u,d}=1$ (dashes).} \label{fig:wgb}
\end{figure}

As a general lesson from this section, we notice a distinctive
features of our 2HDM-III, namely  that the  decay
modes  $W^+ \gamma$  becomes very large,
with a $BR = 2 \times 10^{-1}$ or $10^{-2}$  for some of the
scenarios considered, with $\tan \beta<2$. We also want to compare the general
behavior of these decay modes, within 2HDM-III with the
2HDM-II case. In order to compare the 2HDM-III results with
those in the 2HDM-II, we show in Fig.~\ref{fig:wgb} the  BR$(\dec)_{\rm III} $ in the scenario A and B from 2HDM-III, as well as BR 
$(\dec)_{\rm II} $ from 2HDM-II {\it vs.}
$m_{H^+}$, taking again $\tan \beta =$ 0.3, 0.5, 1, 10. 
We observe that the mode $W^+ \gamma$
is important when 170 GeV $<m_{H^+} <$ 180 GeV and for $0.1 \leq
\tan \beta \leq 1$, taking $\tilde{\chi}_{ij}^{u,d}=-1$. Thus, we find that the effect of
the modified Higgs couplings typical of the 2HDM-III
 shows up clearly in the pattern of charged Higgs
boson decay $\dec$, which can be very different from the 2HDM-II case, and thus
 enrich the possibilities to search for
$H^{\pm}$ states at current (Tevatron) and future (LHC, ILC/CLIC) machines.
\section{\bf Events rates at LHC from the decay $\dec$  }

The production of charged Higgs bosons at hadron colliders  has been evaluated in
 \cite{mapaetal, diaz-sampayo} (also for the Superconducting Super Collider, SSC)
and more recently \cite{newhcprod, DiazCruz:2009ek} (for the LHC) literature, mainly for the 2HDM-II
and its SUSY realization (i.e., the MSSM). In these two scenarios,
when kinematically allowed, the top quark decay
channel $t \to b H^+$ is the dominant $H^\pm$ production mechanism.
Instead, above the threshold for such a decay, the dominant $H^\pm$ production reaction is
gluon-gluon fusion into a 3-body final state, {i.e.}, $gg \to tbH^\pm$\footnote{In fact,
these two mechanisms are intimately related, see below.}. Both processes depend
on the coupling $H^- t\bar b$ and  are therefore sensitive to the
modifications that arise in the 2HDM-III for this vertex. However,
detection of the final state will depend on the charged Higgs boson
decay mode, which could include a complicated final state, that
could in turn be difficult to reconstruct. For these reasons, it is very
important to look for other production channels, which may be easier
to reconstruct. In this regard, the $s$-channel production of charged
Higgs bosons, through the mechanism of $c\bar b$-fusion,  could help to make
more viable the detection of several charged Higgs boson decay
channels \cite{He:1998ie}.

\subsection{\bf Evaluation of event rates from direct production of charged Higgs bosons at the LHC}

The $H^{\pm} \bar{q} q^{\prime}$ vertex with large flavor mixing coupling,
that arises in the 2HDM-III, enables the possibility of studying the
production of charged Higgs boson via the $s$-channel production
mechanism, $c\bar{b} \to H^{+}$ + c.c. This process was
discussed first in Ref.~\cite{He:1998ie} and recently by \cite{DiazCruz:2009ek} in the context of the 2HDM-III.

To illustrate the type of charged Higgs signatures that have the
potential to be detectable at the LHC in the 2HDM-III, we show in the Table~\ref{tab:wg1} 
 the event rates of charged Higgs
boson through the channel
$c\bar{b} \to H^+$ + c.c., alongside the corresponding production
cross sections ($\sigma$'s) and relevant BR($\dec$), for a combination of
masses, $\tan\beta$ and specific 2HDM-III parameters amongst those
used in the previous sections (assuming $m_{h^0}= 120$ GeV,
$m_{A^0}=300$ GeV and the mixing angle at $\alpha = \pi /2$
throughout). In particular, we focus on those cases where the
charged Higgs boson mass is above the threshold for $t \to b H^+$,
for two reasons. On the one hand, the scope of the LHC in accessing
$t \to b H^+$ decays has been established in a rather model
independent way. On the other hand, we have dealt at length with the
corresponding BRs in section III. (As default, we also assume an
integrated luminosity of $10^5$ pb$^{-1}$.) In all cases we use the results of our previously work \cite{DiazCruz:2009ek}.

To illustrate these results, let us comment on one case within the
scenario B, because is the case most conservative. From Table \ref{tab:wg1}, we can see that for Scenario B,
with $(\tilde{\chi}_{ij}^{u}=1, \tilde{\chi}_{ij}^{d}=1)$ and
$\tan\beta=0.3$, we have that the ${H^{\pm}}$ is heavier than $m_t -m_b$,
as we take a mass $m_{H^+}=200$ GeV, thus precluding top decay
contributions, so that in this case $\sigma(pp \to  H^+ X)
\approx 2.2 \times 10^{-2}$ pb, and including the  decay  $\dec$  , in the final state we set
42 events. Even for charged Higgs masses as long as  $300$ GeV, we can still obtain 4 events. The most promising   
rate for decay $\dec$, is for $m_{H^+} = 200$ and $\tan \beta =1$, when we get 63 events.

\squeezetable
\begin{table*}[htdp]
\caption{\label{tab:wg1} Summary of LHC event rates for some parameter combinations within Scenarios  B with
   an integrated luminosity of $10^{5}$
pb$^{-1}$, for the signal $\dec$, through the channel $c
\bar b \to H^+$ + c.c.}

\begin{tabular}{|c|c|c|c|c|c|}
\hline $(\tilde{\chi}_{ij}^{u}, \tilde{\chi}_{ij}^{d})$  & $\tan\beta$ & $m_{H^+}$ in GeV &
$\sigma(pp \to H^+ + X)$ in pb & $BR(\dec)$& Nr. Events\\

\hline \multicolumn{1}{|c|}{ (1,1) }& 0.3 &200 & $ 2.1 \times
10^{2}$ &\begin{tabular}{l} $2 \times 10^{-6} $%
\end{tabular}
& \multicolumn{1}{|c|}{$
\begin{tabular}{r}
42
\end{tabular}
$} \\ \hline

\hline \multicolumn{1}{|c|}{ (1,1) }& 0.3 & 300 & $3.1\times
10$ &\begin{tabular}{l} $1.1\times 10^{-6} $%
\end{tabular}
& \multicolumn{1}{|c|}{$
\begin{tabular}{r}
4
\end{tabular}
$} \\ \hline

\hline \multicolumn{1}{|c|}{ (1,1) }& 1 & 200 & $3$ &\begin{tabular}{l} $2.1 \times 10^{-4} $%
\end{tabular}
& \multicolumn{1}{|c|}{$
\begin{tabular}{r}
63
\end{tabular}
$} \\ \hline

\hline \multicolumn{1}{|c|}{ (1,1)}& 1 & 300 & $0.98$ &\begin{tabular}{l} $4.2 \times 10^{-7} $
\end{tabular}
& \multicolumn{1}{|c|}{$
\begin{tabular}{r}
0
\end{tabular}
$} \\ \hline

\hline \multicolumn{1}{|c|}{ (1,1) }& 10 &200 & $1.1$ &\begin{tabular}{l} $9.8 \times 10^{-5} $%
\end{tabular}
& \multicolumn{1}{|c|}{$
\begin{tabular}{r}
10
\end{tabular}
$} \\ \hline

\hline \multicolumn{1}{|c|}{ (1,1) }& 10 & 300 & $5.2 \times
10^{-1} $ &\begin{tabular}{l} $1.1 \times 10^{-6} $%
\end{tabular}
& \multicolumn{1}{|c|}{$
\begin{tabular}{r}
0
\end{tabular}
$} \\ \hline

\end{tabular}
\label{default6}
\end{table*}

\subsection{Indirect production of charged Higgs bosons at the LHC}

We have found that, in some of the 2HDM-III scenarios
envisaged here, light charged Higgs bosons could exist that have
not been excluded by current experimental bounds, chiefly from LEP2
and Tevatron. Their discovery potential should therefore be studied
in view of the upcoming LHC and we shall then turn our attention now to
presenting the corresponding hadro-production cross sections via an
indirect channel, i.e., other than as secondary products in (anti)top
quark decays and via $c\bar b$-fusion, considered previously.

In previously work \cite{DiazCruz:2009ek}, we evaluate  the  $q\bar q,gg\to t\bar b H^-$ + c.c.
cross sections. We use HERWIG version 6.510 in default configuration, by
onsetting the subprocess {\tt IPROC~=~3839}, wherein we have
overwritten the default MSSM/2HDM couplings and masses with those
pertaining to the 2HDM-III: see Eqs.~(\ref{coups1})--(\ref{sp}).

Now, let discuss  again the
scenario B. From Table \ref{tab:wg2}, we can see that for this scenario,
with $(\tilde{\chi}_{ij}^{u}=1, \tilde{\chi}_{ij}^{d}=1)$ and
$\tan\beta=0.3$, we have that the ${H^{\pm}}$ is heavier than $m_t -m_b$,
as we take a mass $m_{H^+}=200$ GeV, thus precluding top decay
contributions, so that in this case $\sigma(pp \to t\bar b H^- )  
\approx 25.8 $ pb, while the  signal  $\dec$  give
a number of events of  5. When the mass of the charged Higgs take the value of $300$ GeV, we can obtain only 1 event. One can see that for 
$m_{H^+} = 200$ and $\tan \beta =1$, we get 48 events. However, the
 case  most promising   rate for decay $\dec$ is when $m_{H^+} = 200$ and $\tan \beta =10$, we get the spectacular 122 events, which could be important for the LHC collider experiments.

Altogether, by comparing the  $q\bar q,gg\to t\bar b H^-$ + c.c.
cross sections herein with, e.g., those of the MSSM in
\cite{Djouadi:2005gj} or the 2HDM in
\cite{Moretti:2002ht,Moretti:2001pp}, it is clear that the 2HDM-III
rates can be very large and thus the discovery potential in ATLAS
and CMS can be substantial, particularly for a very light
$H^\pm$, which may pertain to our 2HDM-III but not the MSSM or
2HDM-II. However, it is only by combining the production rates of
this section with the decay ones of the previous ones that actual
event numbers at the LHC can be predicted.


\squeezetable
\begin{table*}[htdp]
\caption{\label{tab:wg2} Summary of LHC event rates for some parameter combinations within Scenarios B with
  for an integrated luminosity of $10^{5}$
pb$^{-1}$, for $\dec$ signature, through the channel $q
\bar q,gg \to \bar{t} b H^+$ + c.c.}

\begin{tabular}{|c|c|c|c|c|c|}
\hline $(\tilde{\chi}_{ij}^{u}, \tilde{\chi}_{ij}^{d})$  & $\tan\beta$ & $m_{H^+}$ in GeV &
$\sigma(pp \to H^+ \bar{t} b)$ in pb & BR$(\dec)$ & Nr. Events\\

\hline \multicolumn{1}{|c|}{ (1,1)  }& 0.3 &200 &
25.8 &\begin{tabular}{l}  $2 \times 10^{-6} $%
\end{tabular}
& \multicolumn{1}{|c|}{$
\begin{tabular}{r}
5
\end{tabular}
$} \\ \hline

\hline \multicolumn{1}{|c|}{ (1,1) }& 0.3 & 300 &
 10.1 &\begin{tabular}{l} $ 1.1 \times 10^{-6} $%
\end{tabular}
& \multicolumn{1}{|c|}{$
\begin{tabular}{r}
1
\end{tabular}
$} \\ \hline

\hline \multicolumn{1}{|c|}{ (1,1) }& 1 & 200 &
2.3&\begin{tabular}{l} $ 2.1 \times 10^{-4} $%
\end{tabular}
& \multicolumn{1}{|c|}{$
\begin{tabular}{r}
48
\end{tabular}
$} \\ \hline

\hline \multicolumn{1}{|c|}{ (1,1)}& 1 & 300 &
0.79 &\begin{tabular}{l} $4.2 \times
10^{-7} $%
\end{tabular}
& \multicolumn{1}{|c|}{$
\begin{tabular}{r}
0%
\end{tabular}
$} \\ \hline

\hline \multicolumn{1}{|c|}{ (1,1) }& 10 &200 &
12.5 &\begin{tabular}{l} $ 9.8 \times 10^{-5} $%
\end{tabular}
& \multicolumn{1}{|c|}{$
\begin{tabular}{r}
122
\end{tabular}
$} \\ \hline

\hline \multicolumn{1}{|c|}{ (1,1) }& 10 & 300 &
$0.48 $ &\begin{tabular}{l} $1.1 \times 10^{-6} $%
\end{tabular}
& \multicolumn{1}{|c|}{$
\begin{tabular}{r}
0%
\end{tabular}
$} \\ \hline

\end{tabular}
\label{default5}
\end{table*}

\section{Conclusions}

We have discussed the implications of assuming a four-zero Yukawa for the properties of the charged Higgs boson, within the
context of a 2HDM-III. In particular, we have presented a detailed
discussion of the charged Higgs boson couplings to heavy fermions
and the resulting implications for the decay $\dec$, induced at one-loop level. The latter clearly reflect
the different coupling structure of the 2HDM-III, e.g., with respect
to the 2HDM-II, so that one has at disposal more possibilities to
search for $H^{\pm}$ states at current and future colliders, 
enabling us to distinguish between different models of EWSB.
We have then concentrated our analysis to the case of the LHC and
showed that the production rates of charged Higgs bosons at the LHC
is sensitive to the modifications of the Higgs boson couplings. We
have employed  the results of  the $s$-channel production of $H^{\pm}$
through $c\bar{b}$-fusion and the multibody final state induced by
$gg$-fusion and $q\bar q$-annihilation. Finally, we have determined
the number of events for the most promising LHC signatures for the decay
$\dec$ within 2HDM-III, for both $c \bar{b}\to H^+$ + c.c.
and $q\bar q\to \bar t bH^+$ + c.c. scatterings (the latter
affording larger rates than the former). Armed with these results,
we are now in a position to carry out a detailed study of signal and
background rates, in order to determine the precise detectability
level of each signature. However, this is beyond the scope of
present work and will be the subject of a future publication.

\bigskip

\section*{Acknowledgements}
 We thank to Lorenzo D\'iaz-Cruz and Alfonso Rosado for useful discussions. 
 This work was supported in part by CONACyT and SNI
(M\'exico). J.H-S. thanks in particular CONACyT (M\'exico) for the
grant J50027-F and SEP (M\'exico) by the grant PROMEP/103.5/08/1640.
R.N-P. acknowledges the Institute of Physics BUAP for a warm
hospitality and also financial support by PROMEP-SEP. 

\end{document}